\documentclass[final,3p,times,twocolumn]{elsarticle}
 \biboptions{comma,sort&compress}
\usepackage{here}
 \usepackage{graphicx}
  \usepackage{epsfig}
\usepackage{epsf,amsmath,amssymb,graphicx,dcolumn}
\usepackage{scalefnt,subfigure,ulem,pstricks,cite}

\def\nin{\noindent}
\def\beq{\begin{equation}}
\def\eeq{\end{equation}}
\def\bal#1\eal{\begin{align}#1\end{align}}
\def\bea{\begin{eqnarray}}
\def\eea{\end{eqnarray}}

\newcommand{\pdf}{{\abbrev PDF}}
\newcommand{\qcd}{{\abbrev QCD}}

\newcommand{\abbrev}{\scalefont{.9}}

\newcommand{\eqn}[1]{Eq.\,(\ref{#1})}
\newcommand{\fig}[1]{Fig.\,\ref{#1}}

\newcommand{\sct}[1]{Sect.\,\ref{#1}}

\newcommand{\lhc}{{\abbrev LHC}}
\newcommand{\sm}{{\abbrev SM}}
\newcommand{\mssm}{{\abbrev MSSM}}
\newcommand{\susy}{{\abbrev SUSY}}
\newcommand{\bsm}{{\abbrev BSM}}

\newcommand{\lo}{\text{\abbrev LO}}
\newcommand{\nlo}{\text{\abbrev NLO}}
\newcommand{\nnlo}{\text{\abbrev NNLO}}

\newcommand{\msbar}{\overline{\mbox{\abbrev MS}}}

\newcommand{\reference}[1]{Ref.\,\citep{#1}}

\journal{Nuc. Phys. (Proc. Suppl.)}

\begin{document}

\begin{frontmatter}



\title{Differential Higgs$+$jet production in bottom quark annihilation and gluon fusion}

  \address[label1]{Fachbereich C, Bergische Universit\"at Wuppertal, 42097 Wuppertal, Germany}

 \author[label1,label2]{Marius Wiesemann}
  \address[label2]{TH Division, Physics Department, CERN CH-1211 Geneva 23, Switzerland}
\ead{marius.wiesemann@cern.ch}


\begin{abstract}
\noindent
We present recent developments concerning Higgs production in bottom quark annihilation and gluon fusion. For bottom quark annihilation, we show the transverse momentum distribution of the associated jets. Furthermore, we discuss the distribution of events into $n$-jet bins for $n$=0 and $n>0$ at \nnlo{} and \nlo{}, respectively. For gluon fusion, the quality of the heavy-top limit for differential quantities at $\mathcal{O}(\alpha_s^4)$ is studied by taking into account higher order terms in the $1/m_{\text{top}}$ expansion.
\end{abstract}


\end{frontmatter}


\section{Introduction} 
\nin
After the recent discovery of a scalar particle \citep{Collaboration2012,ATLAS2012} which may turn out to be the Higgs Boson the task is now to explore its properties in detail, in particular its fermionic and bosonic couplings. This is crucial in order to identify it as a Standard Model (\sm{}) Higgs Boson or one embedded in a theory beyond Standard Model (\bsm{}) e.g. a supersymmetric (\susy{}) Higgs Boson. For this purpose precision predictions of the production and decay rate are needed from the theoretical side \citep{Dittmaier2011}.

In the \sm{} the most important production mechanism is gluon fusion, where the Higgs-gluon coupling is mediated through a top quark loop. Due to the complicated loop structure at leading order (\lo{}), higher order correction are usually performed in the so-called heavy-top limit, where the top quark is assumed to be infinitely heavy. For the inclusive cross section \citep{Harlander2002,Anastasiou2002,Ravindran2003} the heavy-top limit has been validated to better than one percent \citep{Harlander2009,Harlander2010,Pak2010}, while for differential observables only rather few studies aim to validate the heavy-top approximation \citep{DelDuca2001a,Alwall2012,Bagnaschi2012,Mantler2012}.

In \susy{} extensions, e.g. the minimal supersymmetric standard model (\mssm{}), the associated production of a Higgs and bottom quarks can supersede gluon fusion for large $\tan\beta$. Two approaches to calculate its cross section have been pursued in the literature. In the so-called four-flavor scheme ($4$FS) one assumes the proton to contain only four light quark flavors and the gluon \citep{Dittmaier2004,Dawson2004,Gao2005}, while in the five-flavor scheme ($5$FS) one allows parton distribution functions (\pdf{}s) for the bottom quark in addition \citep{Maltoni2003,Harlander2003}. In the $4$FS, the infrared divergencies of the final state bottom quarks are regulated by  a finite bottom mass $m_b$. This leads to potentially large logarithms $\ln(m_b/m_H)$. These logarithms are formally resummed into the bottom \pdf{}s in the $5$FS.

In the first part of this talk we will present some recent developments concerning Higgs$+n$-jet production in bottom quark annihilation in the $5$FS, where we will focus on aspects specific to the associated jets. In the final part of this talk we will discuss the validation of the heavy-top limit for differential quantities in gluon fusion.

\section{Higgs+$n$-jet production in bottom quark annihilation}
\label{sct2}
\begin{figure}[t] 
\centering
\subfigure[]{\includegraphics[width=1.8cm]{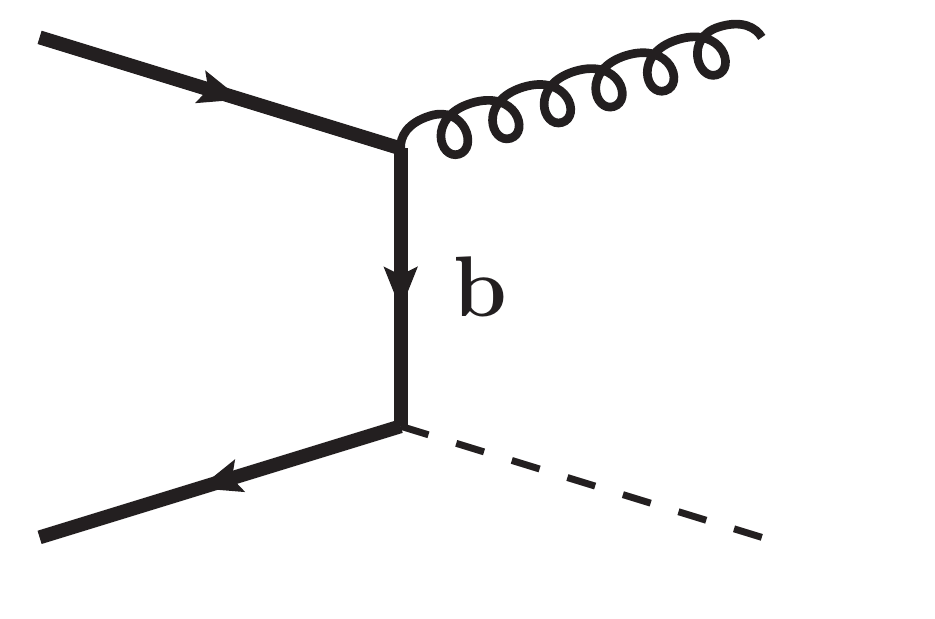}\label{diags_bb1}
}
\subfigure[]{\includegraphics[width=1.8cm]{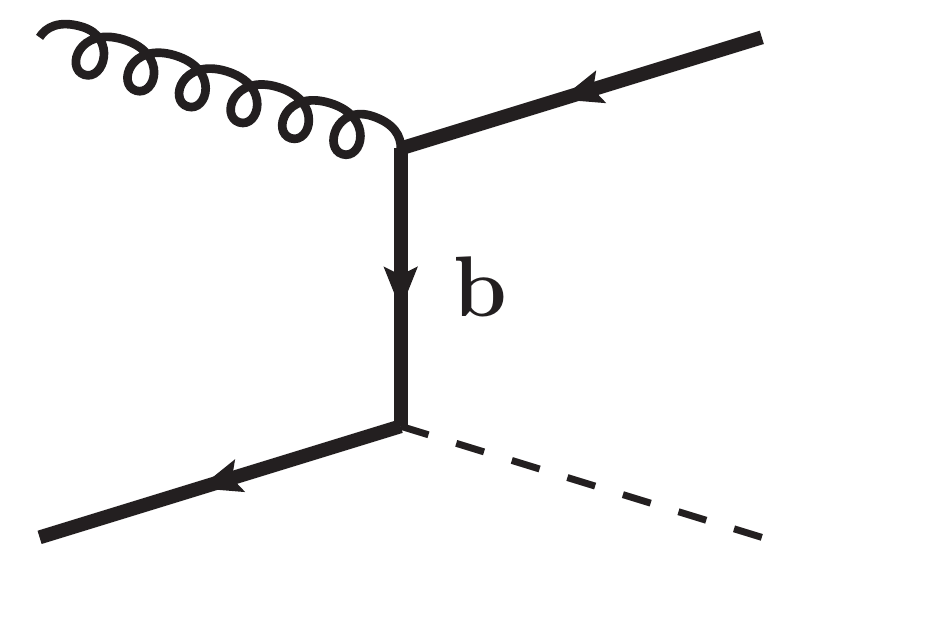}\label{diags_bb2}
}
\subfigure[]{\includegraphics[width=1.8cm]{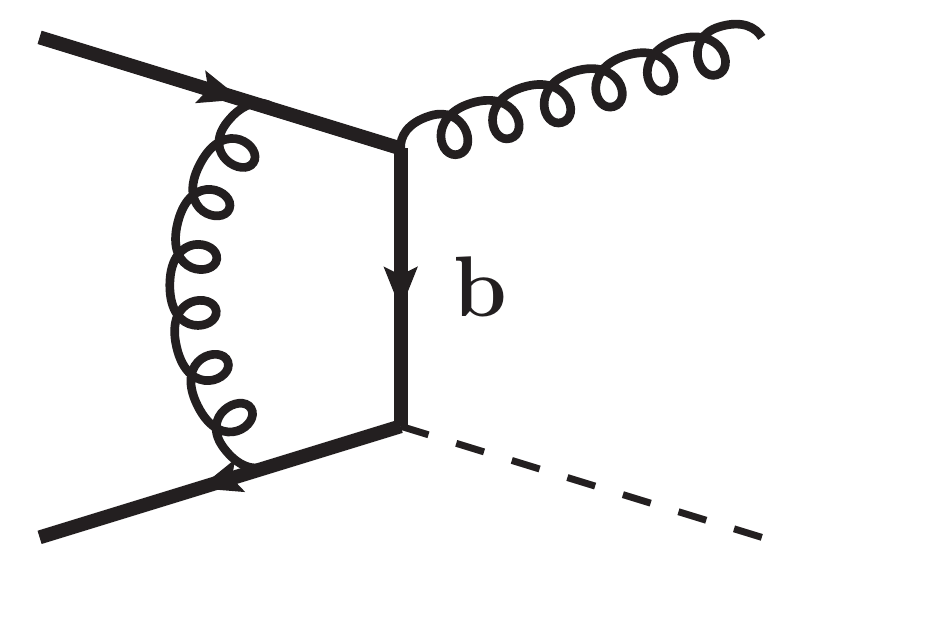}\label{diags_bb3}
}
\subfigure[]{\includegraphics[width=1.8cm]{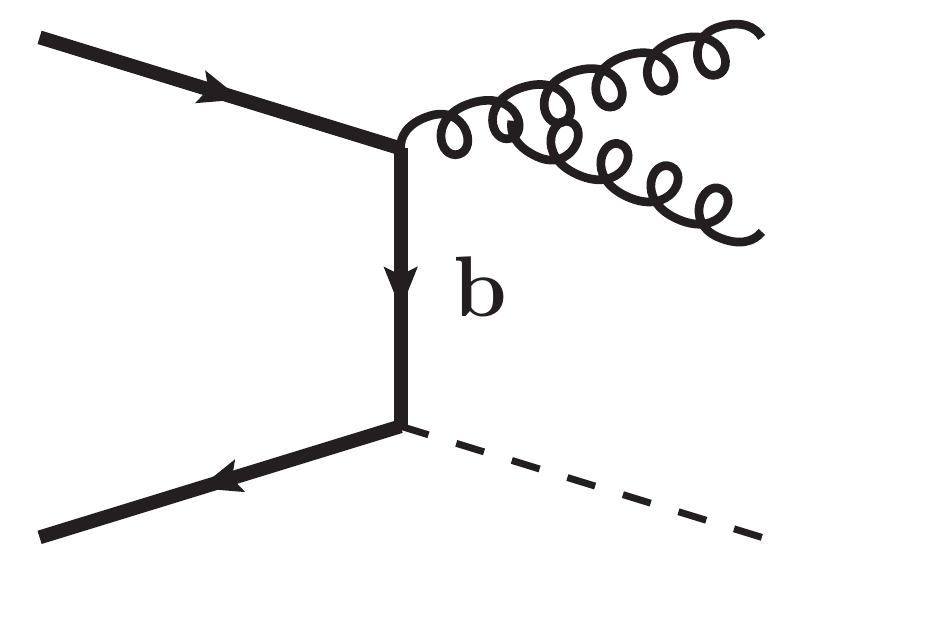}\label{diags_bb4}
}
\caption{\scriptsize The two \lo{} diagrams (a) $b\bar{b}\rightarrow gH$, (b) $gb\rightarrow bH$ and two representative diagrams at \nlo{} for (c) virtual corrections and (d) real emission.}
\label{diags_gg}
\end{figure}

\nin
Since in the \mssm{} bottom quark-associated Higgs production may actually be the dominant mechanism for Higgs production, the studies done for H+jet production in gluon fusion should be supplemented by the bottom-annihilation contribution. Therefore we built a fully differential Monte Carlo integrator for Higgs$+$jet production in bottom quark annihilation through \nlo{} \citep{Harlander2010a,Harlander2012b} using the well known dipole subtraction method \citep{Catani1996}. The corresponding \lo{} diagrams are shown in Fig.\,\ref{diags_bb1}\,and\,\ref{diags_bb2}. Two representative \nlo{} diagrams are given in Fig.\,\ref{diags_bb3}\,and\,\ref{diags_bb4}.
Several checks on the results have been performed: The so-called $\alpha$-parameter~\citep{Nagy1998,Nagy2003} has been implemented to verify the consistency of the dipole subtraction procedure. 
We compared the transverse momentum and rapidity distribution of the Higgs to \reference{Ozeren2010}; observables specific to identified bottom quarks have been cross checked with {\tt MCFM}~\citep{Campbell2002,Campbell}; and a complete comparison of quantities including jets has been performed with the fully automated program {\tt aMC@NLO}~\citep{Frederix2009,Hirschi2011,Frixione2002,Frixione2010}.

We present results for the \lhc{} at $7$ TeV with a Higgs mass of $120$ GeV. The central factorization- ($\mu_F$) and renormalization-scale ($\mu_R$) is $\mu_0 \equiv m_H/4$. Furthermore, all numbers are produced with the MSTW2008 \citep{Martin2009} PDFs and the corresponding strong coupling constant. We insert the bottom quark mass of the $b\bar{b}H$ Yukawa coupling in the $\msbar{}$-scheme at
the scale $\mu_R$, derived from $m_b(m_b) = 4.2$ GeV. Jets are defined using the anti-$k_T$ algorithm with jet radius $R = 0.4$, $p_T^{\text{jet}} > 20$ GeV and $|y^{\text{jet}}| < 4.8$. The results are given in the \sm{}, but according to the studies of \citep{Dittmaier2006,Dawson2011}, they are applicable to the \mssm{} by simply rescaling the $b\bar{b}H$ coupling. 

\begin{figure}[b] 
\centering
\subfigure[]{\includegraphics[width=3.85cm]{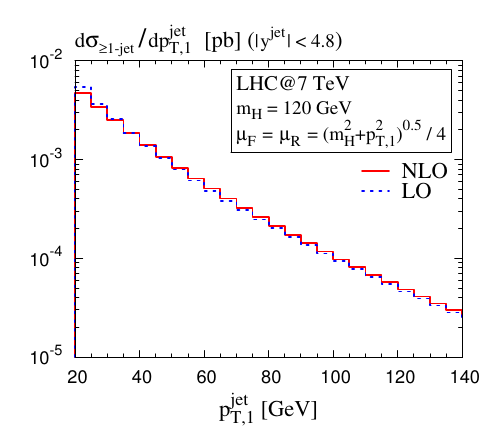}}
\subfigure[]{\includegraphics[width=3.85cm]{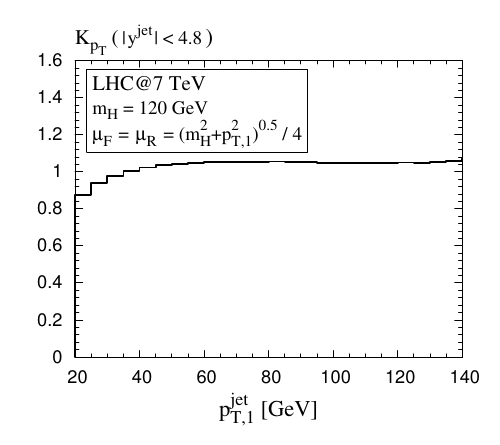}}
\caption{\scriptsize (a) Transverse momentum distribution of the hardest jet and (b) the corresponding $K$-factor. From \reference{Harlander2012b}.}
\label{tmd} 
\end{figure} 

In \fig{tmd} we show the transverse momentum distribution of the hardest jet and the corresponding $K$-factor. To account for effects of high-$p_T$ jets, the central scale choice is modified to be
\bal
\mu_F=\mu_R = \frac14\,\sqrt{m_H^2+(p_{T,1}^{\text{jet}})^2}.
\eal
For this scale choice we observe rather small perturbative corrections $<10\%$ and a flat dependence of the $K$-factor once $p_{T,1}^{\text{jet}}$ is larger than $50$ GeV.

With the knowledge of the total cross section \citep{Harlander2003} our machinery is capable of calculating the jet-vetoed (or $H+0$-jet) cross section at \nnlo{} by simply subtracting the jet contributions from the total cross section
\bal
\label{0jet}
\sigma_{\text{jet-veto}}\equiv\sigma_{0-\text{jet}}= \sigma_{\text{tot}}-\sigma_{\ge 1-{\text{jet}}}.
\eal
This allows us to evaluate the distribution of events into $H+n$-jet bins for $n=0,1,2$ at \nnlo{}, \nlo{} and \lo{}, respectively.\footnote{The fully differential cross section for $b\bar{b}\rightarrow H$ through \nnlo{} was obtained recently \citep{Buehler2012}.}

\begin{figure}[hbt] 
\centering
\subfigure[]{\includegraphics[width=3.85cm]{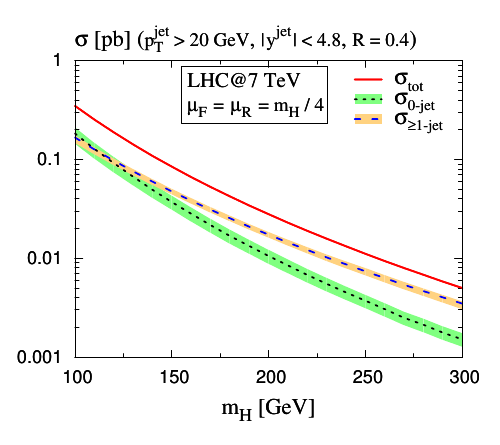}\label{mH1}}
\subfigure[]{\includegraphics[width=3.85cm]{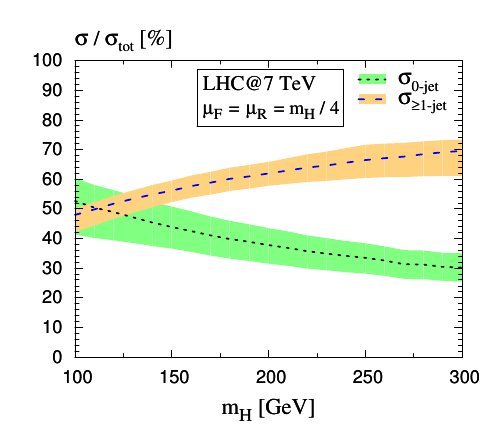}\label{mH2}}
\caption{\scriptsize (a) Higgs mass dependence of the $H + 0$- and $\ge 1$-jet contributions to the total cross section at NNLO and NLO, respectively, from \reference{Harlander2012b}, and (b) the relative contributions normalized by the total cross section.}
\label{mH} 
\end{figure}

In \fig{mH1} the decomposition of the total inclusive cross section (solid, red, no uncertainties) into the exclusive $H+0$-jet (black, dotted) and inclusive $H+$jet (blue, dashed) rate is shown. \fig{mH2} illustrates the relative contributions normalized by the total cross section. Similar to what was found for gluon fusion \citep{catani2001} the contributions including jets increase for higher Higgs masses. The error bands emerge from the quadratically added \pdf{} \citep{Martin2009} and the scale uncertainties. We vary $\mu_R$ and $\mu_F$ around the central scale $\mu_0$ within $[0.5 \mu_0, 2 \mu_0]$.
\nin

\section{Validation of the heavy-top limit for differential Higgs production in gluon fusion}
\nin
In the second part of this talk we investigate the question if the heavy-top approximation of the gluon fusion cross section prediction is an accurate approach for differential quantities. Since higher order corrections, beyond \nlo{}, of the gluon fusion process with the exact top mass dependence are not feasible with current technology, the usual procedure is to calculate the size of perturbative corrections in the heavy-top limit and rescale the cross section including the exact top mass dependence by the obtained $K$-factor. The top mass effects are considered to be small in this procedure. This has been proven to be true to better than $1\%$ for the total inclusive cross section at \nnlo{} \citep{Harlander2010,Pak2010}.

For differential observables \citep{Anastasiou2004,Catani2007}, however, there exist rather few studies concerning the validity of the heavy-top approach \citep{DelDuca2001a,Alwall2012,Bagnaschi2012}. In this section we analyze the quality of the heavy-top limit for distributions at $\mathcal{O}(\alpha_s^4)$ using the calculation of subleading terms in the $1/m_{\text{top}}^2$ expansion for Higgs$+$jet production in gluon fusion through \nlo{} \citep{Harlander2012}. In \fig{diags_gg} some representative diagrams at \lo{} (a), (b) and \nlo{} (c), (d) are shown. The matrix elements were obtained from the authors of \reference{Harlander2009,Harlander2010}. The cancellation of infrared divergencies was achieved using dipole subtraction \citep{Catani1996}.
\begin{figure}[t] 
\centering
\subfigure[]{\includegraphics[width=1.6cm]{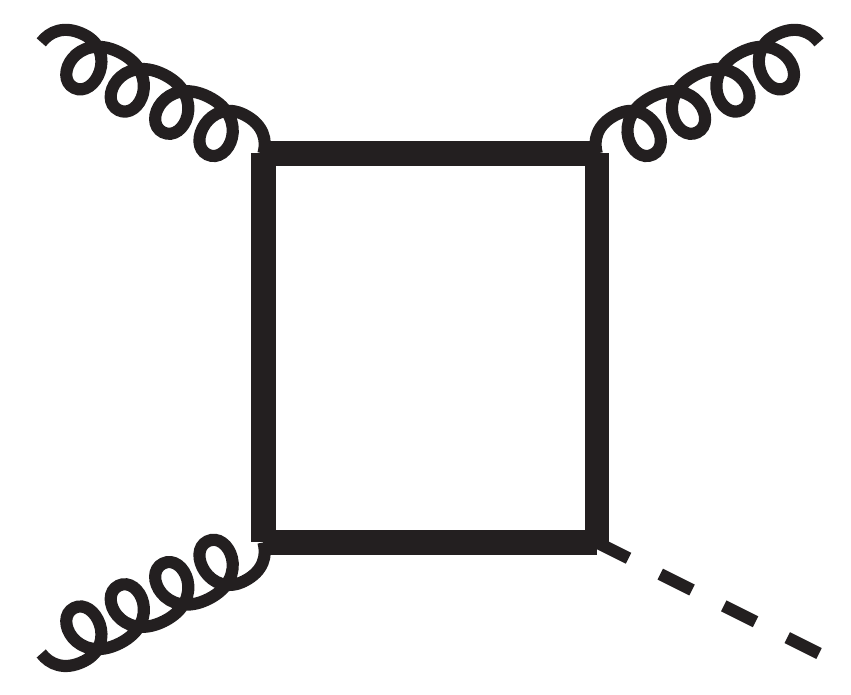}\label{diags_gg1}}\hspace{0.2cm}
\subfigure[]{\includegraphics[width=1.6cm]{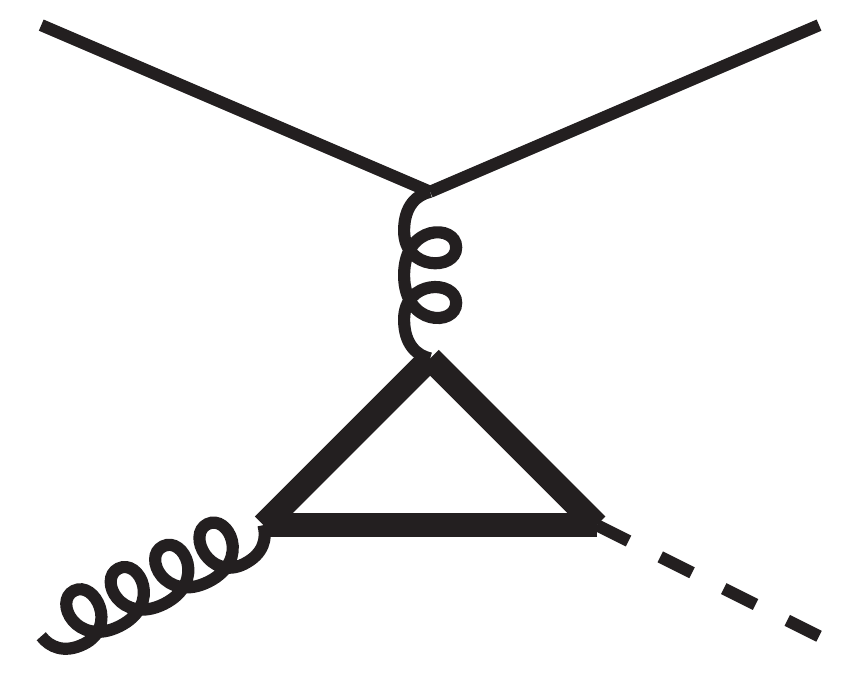}\label{diags_gg2}}\hspace{0.2cm}
\subfigure[]{\includegraphics[width=1.6cm]{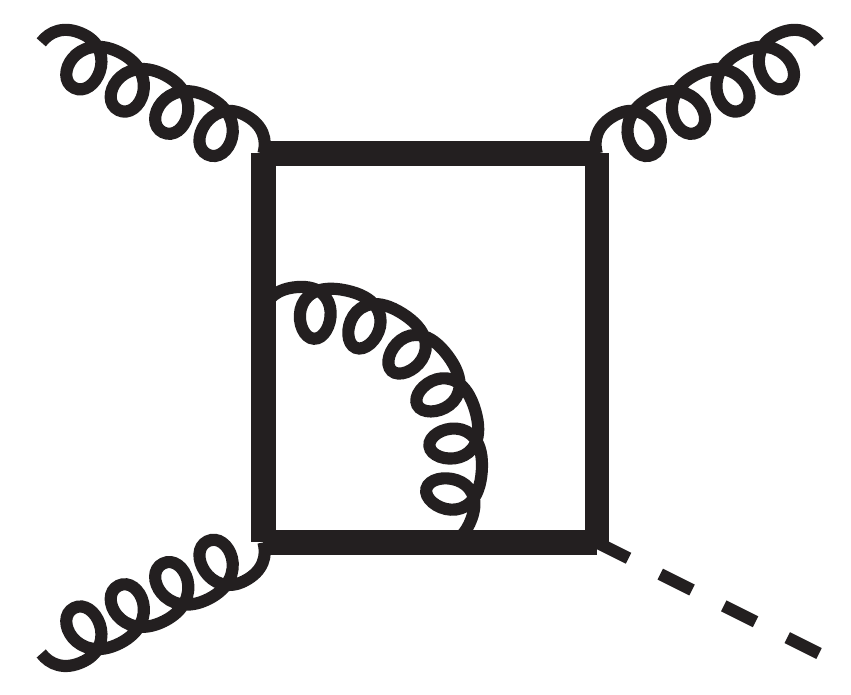}\label{diags_gg3}}\hspace{0.2cm}
\subfigure[]{\includegraphics[width=1.6cm]{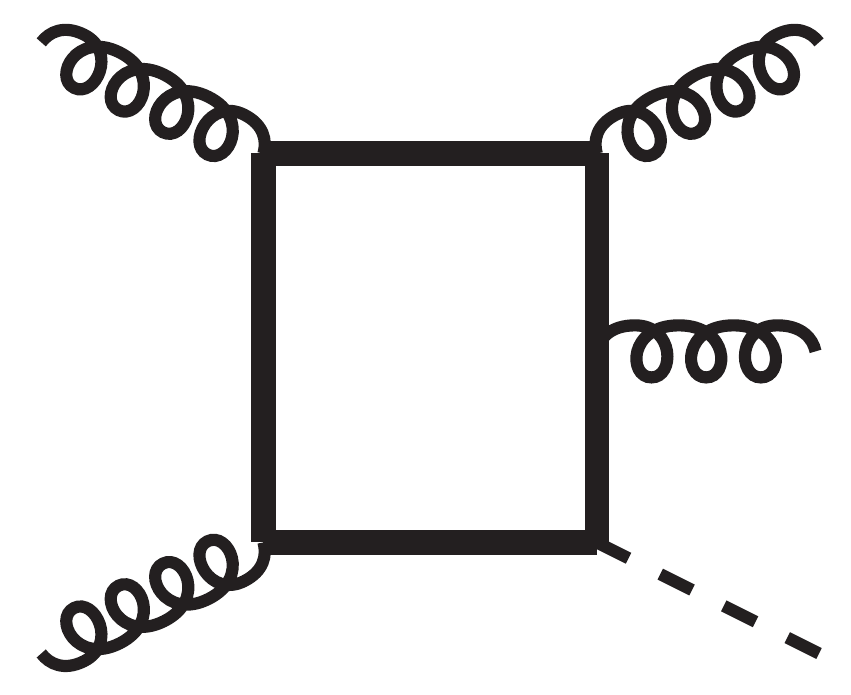}\label{diags_gg4}}
\caption{\scriptsize Representative diagrams for $H+$jet production in gluon fusion at (a), (b) \lo{} and (c), (d) \nlo{}.}
\label{diags_gg}
\end{figure} 
We have performed a number of checks on our results. The most important one is the comparison of the heavy-top limit with the non-resummed part of {\tt HqT}~\citep{Bozzi2003}
. The proper implementation of the dipoles for both the leading and subleading terms in $1/m_{\text{top}}$ was checked by the $\alpha$-parameter \citep{Nagy1998,Nagy2003}.

We use the input parameters according to \sct{sct2} except for the center of mass energy which we choose to be $14$ TeV. This allows us more general conclusions, since the heavy-top approximation is assumed to work worse for higher center of mass energies. Furthermore, we insert an on-shell top mass of $m_{\text{top}}^{\text{OS}} = 172$ GeV.

\begin{figure}[t] 
\centerline{\includegraphics[width=8.cm]{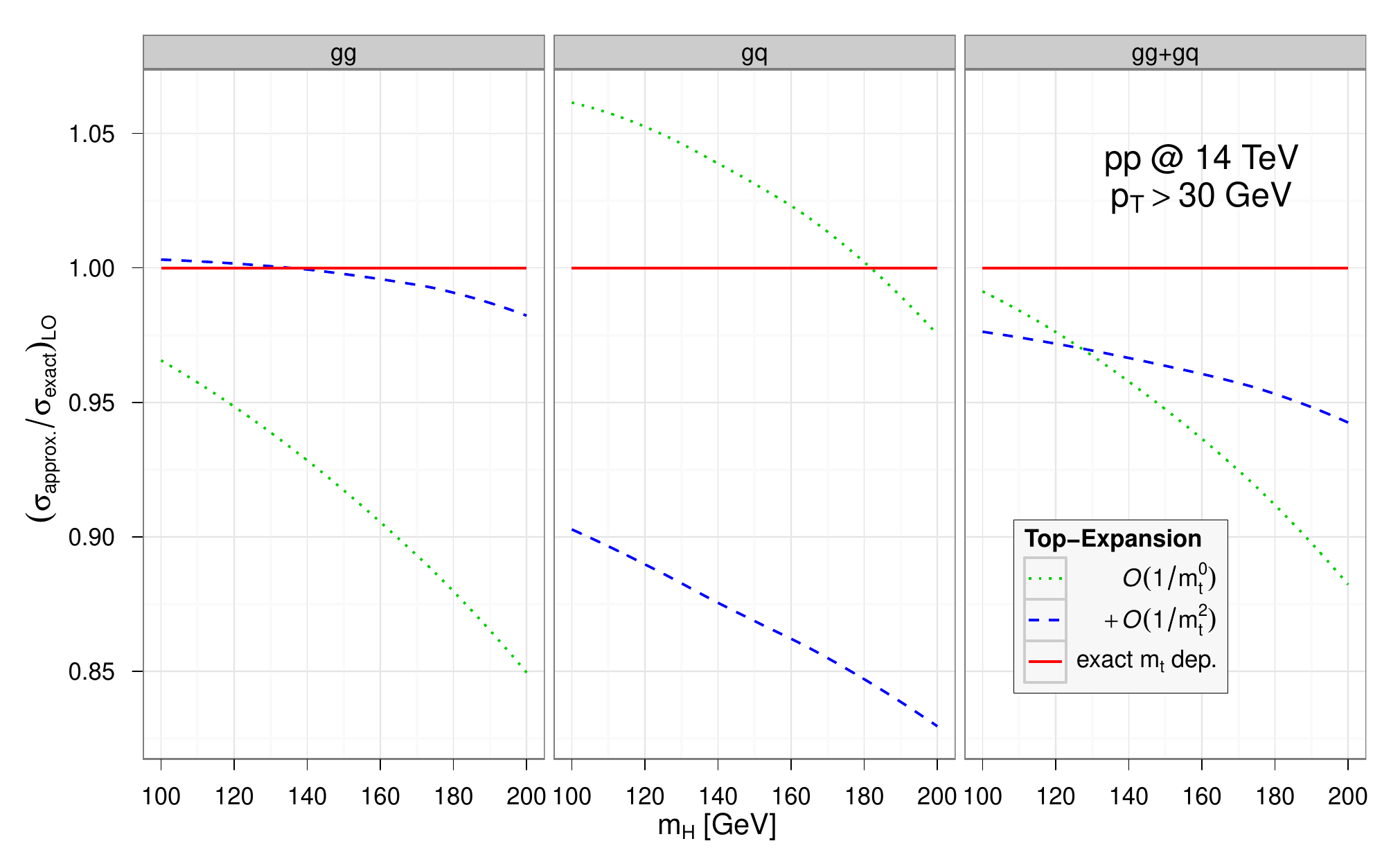}}
\caption{\scriptsize Ratio of the semi-inclusive \lo{} cross section from \eqn{semi-incl} for $p_T^H>30$ GeV when expanded up to $1/m_{\text{top}}^n$ to the exact top mass result, for $n = 0$ (green, dotted) and $n = 2$ (blue, dashed). Left: only $gg$; center: only $qg$; right: sum of $gg$ and $qg$. From \reference{Harlander2012}.}
\label{mHsemi} 
\end{figure}

\fig{mHsemi} shows the semi-inclusive cross section at \lo{} defined with a simple cut on the transverse momentum of the Higgs
\bal
\sigma(p_T^H > p_T^{\text{cut}}) = \int\limits_{p_T^H>p_T^{\text{cut}}} dp_T^H\,\frac{d\sigma}{dp_T^H}
\label{semi-incl}
\eal
as a function of $m_H$, divided into the individual partonic sub-processes $gg$, $gq$ and the sum of both.\footnote{Here and in what follows we omit all $q\bar{q}$ and $qq'$ contributions, since they amount to less than $1\%$ to the cross section.} All cross sections are normalized to the rate with the exact top mass dependence. The dotted green and dashed blue lines denote the normalized cross sections in the pure heavy-top limit and the cross section up to $1/m_{\text{top}}^2$, respectively.
\nin
Looking at the full result in the right plot it is clear that already at \lo{} the heavy-top limit deviates between $1-12\%$ from the exact cross section in the given Higgs mass range and about $2-3\%$ for a Higgs mass of $125$ GeV. This substantiates the importance of reweighting the \lo{} rate including the exact top mass dependence to reduce the uncertainty due to the missing top mass effects.
\begin{figure}[b] 
\centerline{\includegraphics[width=8.cm]{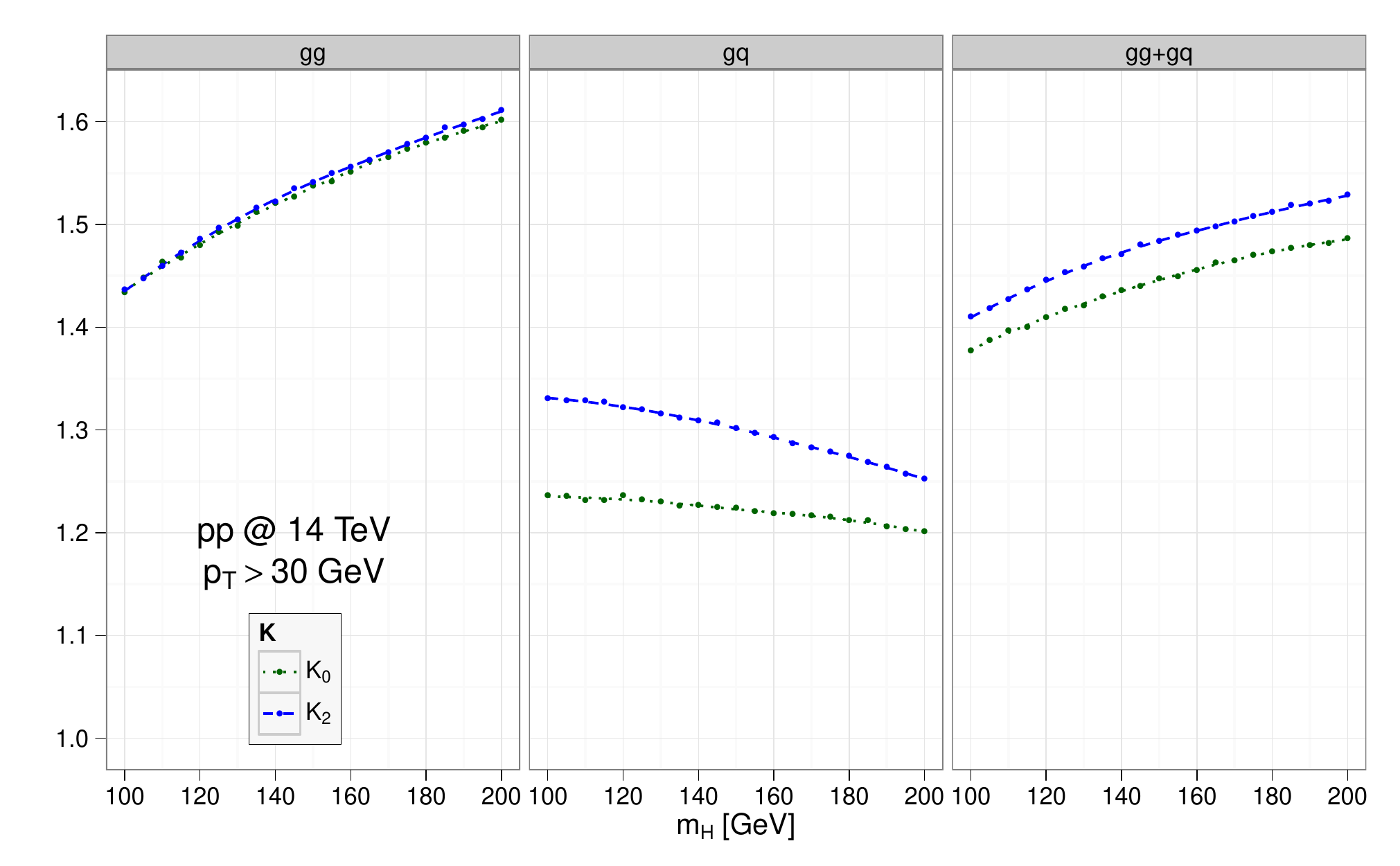}}
\caption{\scriptsize $K$-factors of the semi-inclusive cross section in the pure heavy-top approximation $K_0$ and including subleading $1/m_{\text{top}}^2$ terms $K_2$, separately for the channels $gg$, $gq$ and their sum. The dots show the results of our calculation; the lines have been introduced to guide the eye. From \reference{Harlander2012}.}
\label{semiK} 
\end{figure}
\nin

In \fig{semiK} we examine the $K$-factor $K = \sigma^{\nlo{}} / \sigma^{\lo{}}$ of the semi-inclusive cross section in \eqn{semi-incl} as a function of $m_H$ for the individual channels, where $K_0$ denotes the $K$-factor in the heavy-top limit (dotted green) and $K_2$ includes all terms up to $1/m_{\text{top}}^2$ (dashed blue). The agreement between $K_0$ and $K_2$ for the $gg$ channel is truly remarkable while for the $gq$ channel we find a difference of $5-10\%$. Due to the numerical dominance of $gg$, the overall agreement between $K_0$ and $K_2$ is around $3\%$.

\begin{figure}[t] 
\centerline{\includegraphics[width=8.cm]{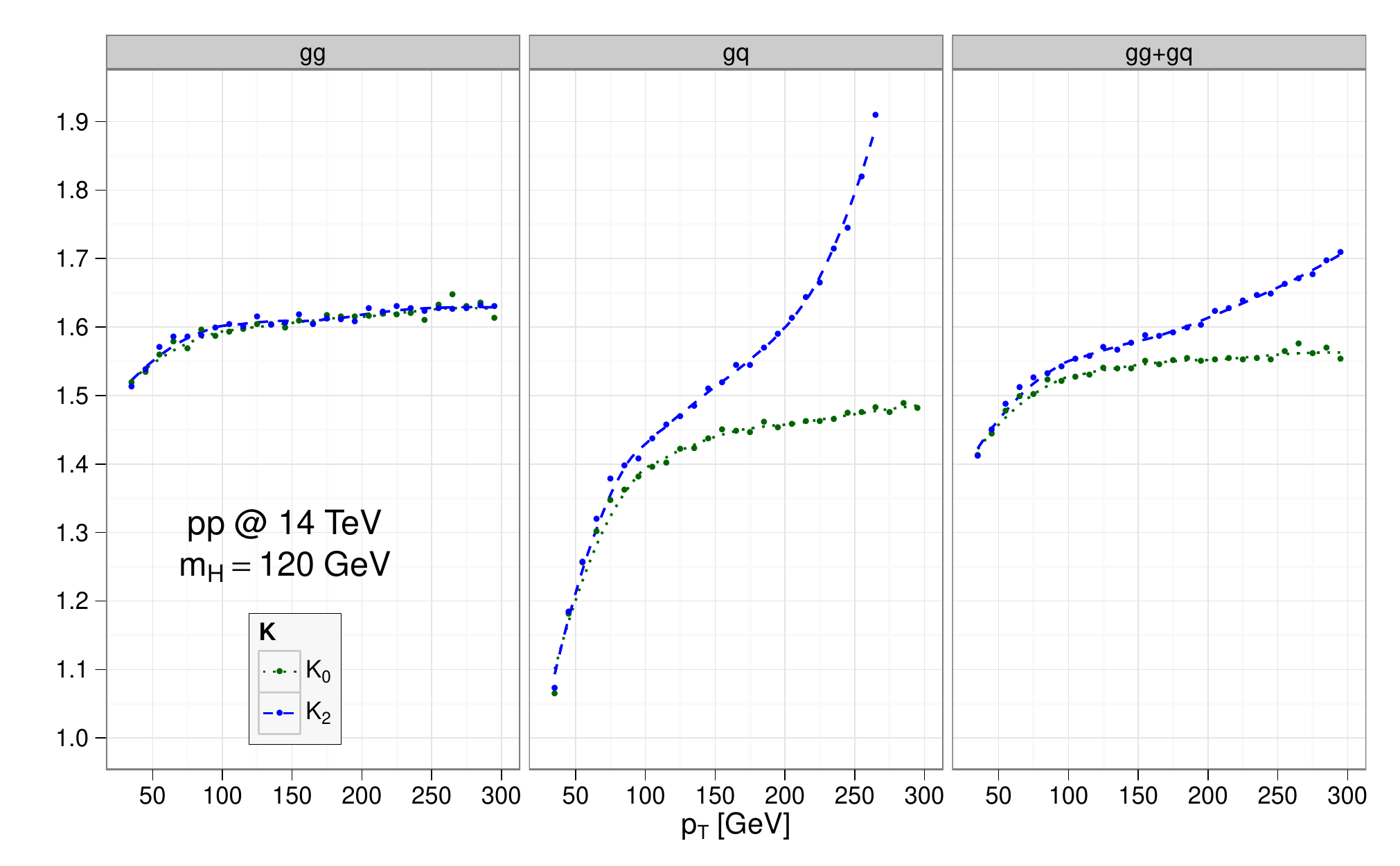}}
\caption{\scriptsize Same as \fig{semiK}, but for the differential cross section $d\sigma / dp_T^H$. From \reference{Harlander2012}.}
\label{pTK} 
\end{figure}

Considering a more differential quantity we show the $p_T^H$-dependent $K$-factors in \fig{pTK}. The observations are similar to what we found for the semi-inclusive cross section: the $K$-factors including leading and subleading mass terms, $K_0$ and $K_2$, are almost identical in the $gg$ channel. For the $gq$ channel, however, the agreement gets lost once $p_T^H > 150$ GeV. In the sum of both channels, the difference remains below $3\%$ for $p_T^H < 150$ GeV, and reaches $10\%$ at $p_T^H = 300$ GeV.
\nin
\section{Conclusions}
\nin
The transverse momentum distribution of the hardest jet at \nlo{} and the individual contributions of $H + n$-jet rates for $n = 0$ (jet-veto) and $n \ge 1$ to the total inclusive cross section for Higgs production in bottom quark annihilation have been presented to \nnlo{} and \nlo{} accuracy.

In the second part of this talk we analyzed the validity of the heavy-top approach for differential quantities in gluon fusion. The behaviour of the $K$-factors including subleading top mass terms with respect to the pure heavy-top $K$-factors suggests that, for quantities which are integrated over $p_T^H$ and the $p_T^H$-distribution for $p_T^H<150$ GeV, the \qcd{} corrections can be safely calculated in the heavy-top limit; the accuracy remains within $2-3\%$. The best prediction, however, should be calculated at \lo{} using the full top-mass dependence, and then reweighted by these QCD corrections.
\section*{Acknowledgements}
\nin
I would like to thank the TH department of CERN for kind hospitality. I am indebted to Stefano Frixione, Robert Harlander, Tobias Neumann and Kemal Ozeren for fruitful discussion and collaboration. This work was supported by {\abbrev ERC}, grant 291377 "LHCTheory" and {\abbrev BMBF}, contracts 05H09PXE and 05H12PXE.










\bibliographystyle{elsarticle-mod}
\bibliography{Proceedings}

\end{document}